\documentclass[final,12pt,3p]{elsarticle}
\usepackage{latexsym}
\usepackage{amssymb}
\usepackage{amsmath}
\usepackage{cancel}
\usepackage{color}
\definecolor{red}{rgb}{1.0,0.0,0.0}
\definecolor{blue}{rgb}{0.0,0.0,1}
\definecolor{green}{rgb}{0.29, 0.33, 0.13}
\newcommand{\ket}[1]{\left| #1 \right>} 
\newcommand{\bra}[1]{\left< #1 \right|} 
\newcommand{\meanv}[1]{\left< #1 \vphantom{#1} \right>} 
 
\newcommand{\tr}{\mbox{tr}}
\begin{document}
\begin{frontmatter}
\title{A comparative study on different non-Hermitian approaches for modeling open quantum systems}
\author[label1]{Santiago Echeverri-Arteaga}
\author[label1]{Herbert Vinck-Posada}
\address[label1]{Departamento de F\'isica, Universidad Nacional de Colombia, 111321, Bogot\'a, Colombia}
\author[label2]{Edgar A. G\'omez\corref{cor1}}
\address[label2]{Programa de F\'isica, Universidad del Quind\'io, 630004, Armenia, Colombia\fnref{label4}}
\cortext[cor1]{Corresponding author}
\ead{eagomez@uniquindio.edu.co}
%
\begin{abstract}
In this paper, we test and compare the performance of two different theoretical methods that use anti-Hermitian terms for modeling open quantum systems. It is, the non-Hermitian quantum mechanics method (NHQM) and the recent methodology known as corrected non-Hermitian effective Hamiltonian (corrected NHEH) approach. In particular, we provide a comprehensive analysis of the performance of these two theoretical approaches by describing the dynamics of the pumped-dissipative Jaynes--Cummings model. Our results demonstrate that whereas the NHQM method fails for describing correctly the time evolution of the density matrix elements as well as the quantum fidelity measure. The corrected NHEH approach shows an excellent agreement with simulations based on the Lindblad master equation formalism.
\end{abstract}
\end{frontmatter}
%
\section{Introduction}\label{intro}
\noindent
Over the last decades, the theory of open quantum systems has gained considerable attention due to their potential for describing dissipation, decoherence and irreversible processes~\cite{Petruccione:2002,Carmichael:1991,Schlosshauer:2008}. Within this framework, there have been different proposals for studying the interaction of a quantum system coupled to the environment, and particularly theoretical descriptions based on non-Hermitian approaches have been considered exclusively for describing dissipation. Two of the most popular methods for incorporating dissipation in the quantum theory are the non-Hermitian quantum mechanics (NHQM) method and the non-Hermitian effective Hamiltonian (NHEH) approach. The first theoretical approach has been introduced in the literature for modeling the effect of dissipation in quantum systems, and it has been applied for studying a two-level system coupled to a generic dissipative environment~\cite{Sergi:2013}. This method has also been used to study both comparison and unification of non-Hermitian and Lindblad approaches with possible applications to open quantum systems~\cite{Zloshchastiev:2014}. In recent years, some achievements have been made within the NHQM method, for example, the development of an interesting formalism for studying multi-time correlation functions for systems whose dynamics are governed by anti-Hermitian Hamiltonians~\cite{Sergi:2015}, and investigations on the onset of the disorder in quantum dissipative systems and quantum entropy~\cite{Sergi:2016}. This approach has also been applied to studies related to phenomena associated with the wave propagation, dissipation in dielectric media and plasmonic waveguides~\cite{Zloshchastiev:2016a,Zloshchastiev:2016b}. Moreover, recently it has been considered in studies of energy transfer in quantum photobiological complexes~\cite{Zloshchastiev:2017}.
The second theoretical approach has been applied in the past to investigate shell models in atomic and nuclear physics~\cite{Volya:2003,Volya:2014} as well as in studies related with molecular physics and quantum chemistry~\cite{Pavlov-Verevkin:1988,Desouter-Lecomte:1999}. It has also been applied to the domains of many-body quantum systems made of ultracold atoms loaded into periodic lattice structures~\cite{Wimberger:2013,Mazzucchi:2016}, as well as in theoretical studies of quantum driven-dissipative systems~\cite{Minganti:2016}, critical behavior and phase transitions in quantum many-body systems~\cite{Ashida:2016,Florentin:2014}. Moreover, it has been shown that this theoretical approach can be adapted to both Markovian and non--Markovian master equations~\cite{Yi:2001,Huang:2008}, and in dissipative quantum systems in which the adiabatic approximation can be carried out~\cite{Yi:2007}. Despite the growing interest in the use of non-Hermitian approaches, important issues regarding the quantum dynamics originated by these methodologies have not been dealt with in depth. For example, the sensitivity to initial conditions and
the impact of different types of environments on the dynamics of an open system, the feasibility of incorporating irreversible processes providing gain in the system through anti-Hermitian terms in the NHQM method or modeling of nonlinear irreversible processes with anti-Hermitian terms in the NHEH approach. Very recently, theoretical works devoted to understanding the reliability of the NHEH approach have conducted a quantitative analysis demonstrating the poor performance of this approach in the description of the time evolution of density matrix elements, observables and the steady-state of a quantum system. As a consequence of these studies, a corrected version of the NHEH approach has been proposed~\cite{Echeverri:2018} and successfully applied to the modeling of Jaynes--Cummings model combined with pump and dissipation mechanisms~\cite{Echeverri:2018b}.
The aim of this work is twofold: on the one hand, to examine and compare the performance between the NHEH approach and the NHQM method for describing dissipation in a well-known open quantum system of the cavity quantum electrodynamics (cQED). On the other hand, to investigate for the first time the performance NHQM method when are included in the system dissipation and pumping through anti-Hermitian terms. This paper is organized as follows. In Section~\ref{sec:formalism} we present the model that describes the pumped-dissipative QD--cavity system as well as two different theoretical methods based on anti-Hermitian approaches. More precisely, subsection~\ref{subsec:one} presents a review of basic theory of the NHQM method and its application to the model of the dissipative QD--cavity system. Subsection~\ref{subsec:two} describes briefly the theory related to the corrected NHEH approach as well as its application to the dissipative QD--cavity system. In Section~\ref{Results_discussion} we present and discuss our numerical results which are summarized in Section~\ref{conclusions}.
\section{Theoretical model}\label{sec:formalism}
\noindent
The Jaynes-Cummings (JC) model is widely considered to be the most important theoretical approach in quantum optics for describing the interaction between light and matter. Specifically, this model describes a two-level system (TLS) interacting with an electromagnetic cavity mode that in the rotating wave approximation its Hamiltonian is given by ($\hbar=1$)
\begin{equation}\label{eq:H}
\hat{H}_{JC}=\omega_x\hat{\sigma}^{\dagger}\hat{\sigma}+\omega_c\hat{a}^{\dagger}\hat{a}+g(\hat{\sigma}\hat{a}^{\dagger}+\hat{a}\hat{\sigma}^{\dagger}),
\end{equation}
where $g$ is the light-matter interaction constant between the cavity mode and the TLS. Additionally, $\omega_{x}$ and $\omega_{c}$ are the frequencies associated with the TLS and the cavity mode, respectively. In particular, we denote by $\hat{a}$ and $\hat{\sigma}=\ket{G}\bra{X}$ the annihilation and lowering operators for the cavity mode and the TLS, such that the action of the lowering operator on the excited state $\ket{X}$ leads to the ground state $\ket{G}$. Interestingly, the JC model has a conserved quantity that is 
associated with number of excitations of the system and it is  defined through the operator $\hat{N}=\hat{a}^{\dagger}\hat{a}+\hat{\sigma}^{\dagger}\hat{\sigma}$ that is diagonal in the bare-states basis $\big\{\ket{\alpha,n}\equiv\ket{\alpha}\vert_{\alpha=G}^{X}\otimes\ket{n}\vert_{n=0}^{\infty}\big\}$. In this basis of states, $n$ and $\alpha$ corresponds to the number of photons in the cavity and one of the two possibles states of the TLS, respectively. Taking into account that the operator $\hat{N}$ defines the conserved quantity it is possible to study the quantum dynamics of the system within separate subspaces of the full state-space. In fact, each subspace is characterized by the eigenvalue $n$ of the operator $\hat{N}$ and it is called the $n$th rung in the JC ladder of states. \\
It is well-known that the Hamiltonian system is given by Eq.~(\ref{eq:H}) is far from describing any real physical situation since it is completely integrable~\cite{Scully:1996} and the light remains always inside the cavity. Therefore, irreversible processes must be considered in order to incorporate the influence of the environment on the dynamics of the JC model, and it is commonly done through a master equation within the Lindblad formalism. In this theoretical approach, the master equation that describes the dissipative JC model with three irreversible processes such as the leakage of photons from the cavity at rate $\kappa$, spontaneous emission at the rate $\gamma_{x}$ and linear pumping of cavity photons at the rate $P$ reads
\begin{equation}\label{exactnumEq}
\frac{d\hat{\rho}}{dt}=\mathcal{H}_{\hat{H}_{JC}}(\hat{\rho})+\frac{\kappa}{2}\mathcal{L}_{\hat{a}}(\hat{\rho})+\frac{\gamma_{x}}{2}\mathcal{L}_{\hat{\sigma}}(\hat{\rho})+\frac{P}{2}\mathcal{L}_{\hat{a}^\dagger}(\hat{\rho}).
\end{equation}
Notice that the first term describes the unitary dynamics of the quantum system, whereas the last three terms correspond to the irreversible processes mentioned above, respectively. Additionally, this master equation in the Lindblad form preserves the trace and hermiticity of the density operator $\hat{\rho}$ at any time $t$.
Here we have defined the superoperators $\mathcal{H}_{\hat{X}}(\cdot)=i(\cdot\hat{X}^\dagger-\hat{X}\cdot)$ and $\mathcal{L}_{\hat{X}}(\cdot)=2\hat{X}\cdot\hat{X}^{\dagger}-\hat{X}^{\dagger}\hat{X}\cdot-\cdot\hat{X}^{\dagger}\hat{X}$ for an arbitrary operator $\hat{X}$, and we will refer to this method as the exact calculation for comparison purposes with the other methods presented in this work.
%
%
\subsection{Non-Hermitian quantum mechanics method}\label{subsec:one}
\noindent
The aim of this subsection is to provide the main results within the NHQM formalism in a simple and self-contained manner, moreover, it will be most necessary for what follows. Let us start considering the case of a quantum system which is described by a Hermitian operator that can be partitioned into Hermitian and anti-Hermitian parts
\begin{equation}\label{eq:001}
\hat{H}=\hat{H}_{+}+\hat{H}_{-},
\end{equation}
where $\hat{H}_{\pm}=\pm\hat{H}^{\dagger}_{\pm}$. A straightforward route towards studying the non-Hermitian quantum dynamics is the Schr\"odinger equation
\begin{equation}\label{eq:002}
\frac{\partial}{\partial t}\ket{\psi(t)}=-\frac{i}{\hbar}(\hat{H}_{+}+\hat{H}_{-})\ket{\psi(t)},     
\end{equation}
which can be formulated in terms of the density operator $\hat{\rho}'=\ket{\psi(t)}\bra{\psi(t)}$ as follows:
\begin{equation}\label{eq:003}
\frac{\partial}{\partial t}\hat{\rho}'(t)=\mathcal{H}_{\hat{H}_{+}}\big(\hat{\rho}'(t)\big)+\mathcal{H}_{\hat{H}_{-}}\big(\hat{\rho}'(t)\big).
\end{equation}
%
A fact well-known in non-Hermitian approaches is that the quantum dynamics is not unitary, and as a consequence of this, the trace of the density operator is not preserved in general. The NHQM formalism surmount this issue by introducing the normalized density operator $\hat{\rho}=\hat{\rho}'/\tr(\hat{\rho}')$ into the Eq.~(\ref{eq:003}) to ensures the probabilistic interpretation of the conventional quantum mechanics. Thus, the time evolution of the density operator is given by 
\begin{equation}\label{eq:004}
\frac{\partial}{\partial t}\hat{\rho}(t)=\mathcal{H}_{\hat{H}_{+}}\big(\hat{\rho}(t)\big)+\mathcal{H}_{\hat{H}_{-}}\big(\hat{\rho}(t)\big)+2i\hat{\rho}(t)\tr\big(\hat{\rho}(t)\hat{H}_{-}\big), 
\end{equation}
and the expectation value at any later time $t$ of an operator $\hat{\chi}$ can be computed through $\meanv{\chi}_t=\tr\big(\hat{\rho}(t)\hat{\chi}\big)$. It is worth to mention that this theoretical approach involves an interplay between effects induced by the environment and non-linear quantum mechanics as it has already been mentioned in previous works~\cite{Gisin:1981,Gisin:1982}. In order to illustrate this fact as well as the  application of the NHQM formalism to the dissipative case of the JC model, we consider explicitly both the Hermitian and anti-Hermitian terms as follows:
\begin{align}
\hat{H}_{+}&=\hat{H}_{JC},\label{her:op:plus}\\
\hat{H}_{-}&=-i\frac{\kappa}{2}\hat{a}^{\dagger}\hat{a}-i\frac{\gamma_{x}}{2}\hat{\sigma}^{\dagger}\hat{\sigma},\label{her:op:minus}
\end{align}
and taking into account the Eq.~(\ref{eq:004}), it is straightforward to write explicitly the relevant dynamical equations for the first rung in the JC ladder of states. It is 
\begin{align}\label{bloch:eqs}
\dot{\rho}_{G0G0}&=\kappa \rho_{G1G1} \rho_{G0G0}+\gamma_x\rho_{X0X0} \rho_{G0G0},\notag
\\
\dot{\rho}_{G1G1}&=-ig(\rho_{X0G1}-\rho_{G1X0})-\kappa\rho_{G1G1}(1-\rho_{G1G1}) +\gamma_x \rho_{G1G1} \rho_{X0X0},\notag
\\
\dot{\rho}_{X0X0}&=-ig(\rho_{G1X0}-\rho_{X0G1})+\kappa\rho_{G1G1}\rho_{X0X0} -\gamma_x \rho_{X0X0}(1-\rho_{X0X0}) \rho_{X0X0},\notag
\\
\dot{\rho}_{G1G0}&=-i\omega_c\rho_{G1G0}-ig\rho_{X0G0}-\rho_{G1G0}\left(1-2\rho_{G1G1}\right)\frac{\kappa}{2}+\gamma_x \rho_{G1G0}\rho_{X0X0},\notag
\\
\dot{\rho}_{G1X0}&=-ig(\rho_{X0X0}-\rho_{G1G1})-\rho_{G1X0}\Big(i(\omega_c-\omega_x)+(1-2\rho_{G1G1})\frac{\kappa}{2}+(1-2\rho_{X0X0})\frac{\gamma_x}{2}\Big),\notag
\\
\dot{\rho}_{X0G0}&=-i\omega_x\rho_{X0G0}-ig\rho_{G1G0}+\kappa\rho_{X0G0}\rho_{G1G1} -\rho_{X0G0}(1-2\rho_{X0X0})\frac{\gamma_x}{2}.
\end{align}
Notice that we have used the notation $\dot{\rho}_{\alpha n,\beta m}=\bra{\alpha n}d\hat{\rho}(t)/dt\ket{\beta m}$ to indicate  time-derivative of an arbitrary element of the density operator. 
%
\subsection{Non-Hermitian effective Hamiltonian approach}\label{subsec:two}
\noindent
Recently, we have introduced in the literature the corrected NHEH approach~\cite{Echeverri:2018,Echeverri:2018b} as an interesting theoretical method for modeling open quantum systems. This methodology takes advantage of conserved quantities in the system, and in particular, allows to solve quantum dynamics of the system within separate subspaces that belong to the full states-space. Additionally, this method has several key advantages since it works for arbitrary initial conditions and can be adjusted to an iterative approach in such a way that allows correcting the dynamics systematically. In particular and for our purposes,  the Hermitian and anti-Hermitian terms given by the Eqs.~(\ref{her:op:plus}),(\ref{her:op:minus}) 
define a conserved quantity, i.e. $[\hat{H},\hat{N}]=0$
when it is considered the operator $\hat{H}=\hat{H}_{JC}+\hat{H}_{-}$. Therefore, the corrected NHEH approach can be implemented easily by solving the equation:
\begin{align}\label{Fullcorrected}
\frac{d\hat{\rho}(t)}{dt}^{[n_{max}-j]}&=\mathcal{H}_{\hat{H}}(\hat{\rho}(t)^{[n_{max}-j]})+(1-\delta_{0,j})\Big(\frac{\kappa}{2}\mathcal{C}_{\hat{a}}(\hat{\rho}(t)^{[n_{max}-j+1]})+\frac{\gamma_x}{2}\mathcal{C}_{\hat{\sigma}}(\hat{\rho}(t)^{[n_{max}-j+1]})\Big)\notag\\
&+(1-\delta_{j,n_{max}})\frac{P}{2}
\mathcal{C}_{\hat{a}^{\dagger}}(\hat{\rho}(t)^{[n_{max}-j-1]}),
\end{align}
with $j=0,1,2,3,...,n_{max}$. It is important to note that the superscript $[n]$ placed on the density operator means that a set of dynamical equations at the $n$th rung should be solved as a function of time $t$, and $n_{max}$ defines the maximum rung to be corrected. The superoperator $\mathcal{C}_{\hat{O}}(\cdot)=\hat{O}\cdot\hat{O}^\dagger$ has been introduced in the above equation for taking into account the population transfer between adjacent rungs as a function of time. Furthermore, notice that depending on the particular operator $\hat{O}$ the population will be transferred to the upper or lower rungs in the JC ladder of states.
\section{Results and discussion}\label{Results_discussion}
\begin{figure}[ht!]
\centering
\includegraphics{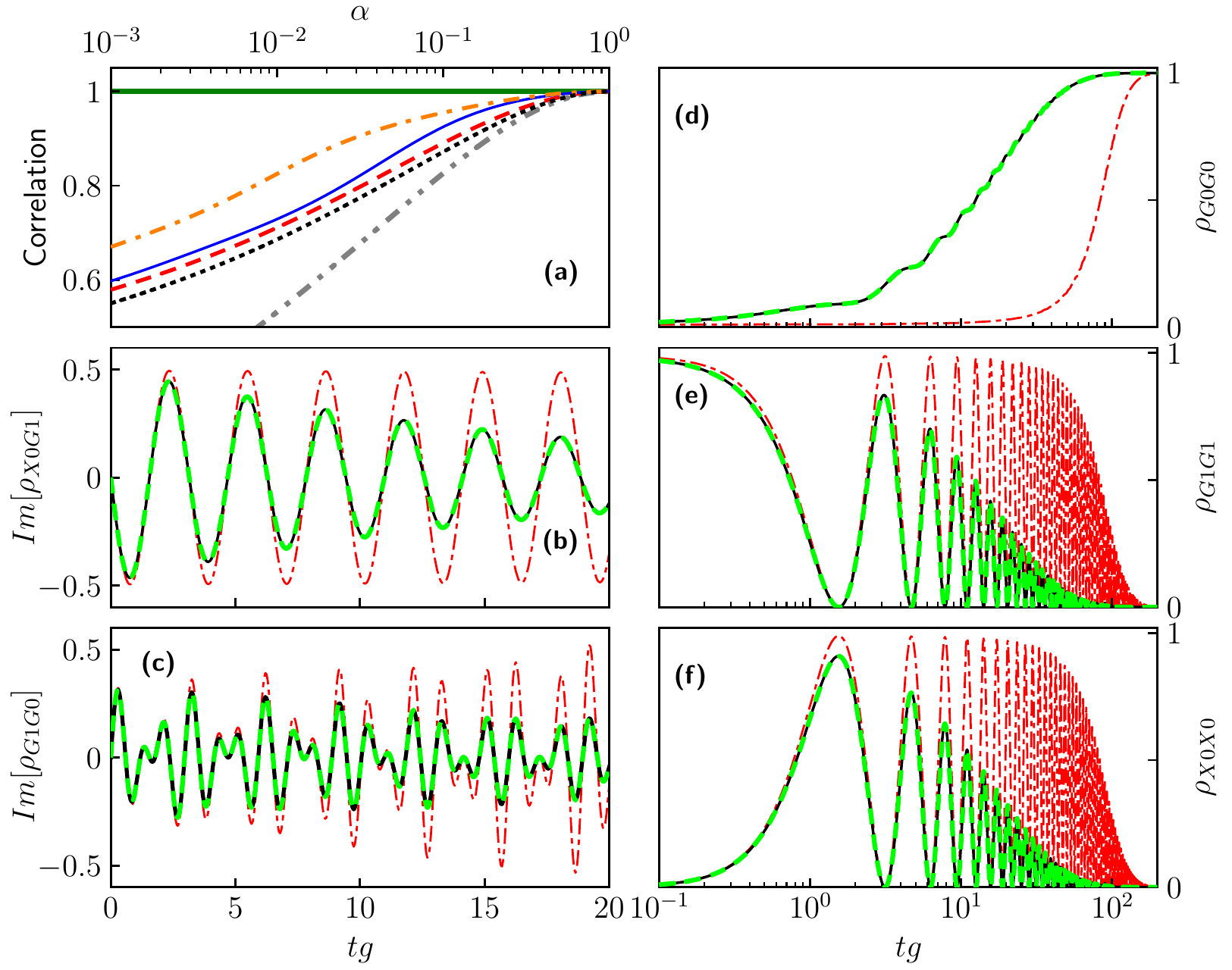}\caption{Panel (a) shows the correlation coefficient as a function of the parameter $\alpha$ (see details in text). The numerical results based on the Lindblad master equation and the NHQM method are shown for populations $\rho_{X0X0}$, $\rho_{G1G1}$ and $\rho_{G0G0}$ with dot-dashed yellow, solid blue and dotted black lines, respectively. Similar numerical calculations have been performed the coherences $\rho_{X0G1}$ and $\rho_{G0G1}$ with dashed red and double dot-dashed gray lines.
The comparison between numerical results based on the Lindblad master equation and the corrected NHEH approaches are shown with a thick solid green line. Panels (b)-(c) show the imaginary part of the coherence $Im[\rho_{X0G1}]$ and $Im[\rho_{G1G0}]$ as a function of time, respectively. Numerical calculations based on NHQM and NHEH approaches are shown with dot-dashed red and dashed green lines, respectively. The exact numerical calculations based on the Lindblad master equation are shown with a solid black line. Panels (d)-(e)-(f) show the same calculations but for the populations $\rho_{G0G0}$, $\rho_{G1G1}$ and $\rho_{X0X0}$, respectively.
(For interpretation of the references to color in this figure legend, the reader is referred to the web version of this article.)}\label{Fig1}
\end{figure}

\begin{figure}[ht!]
\centering
\includegraphics{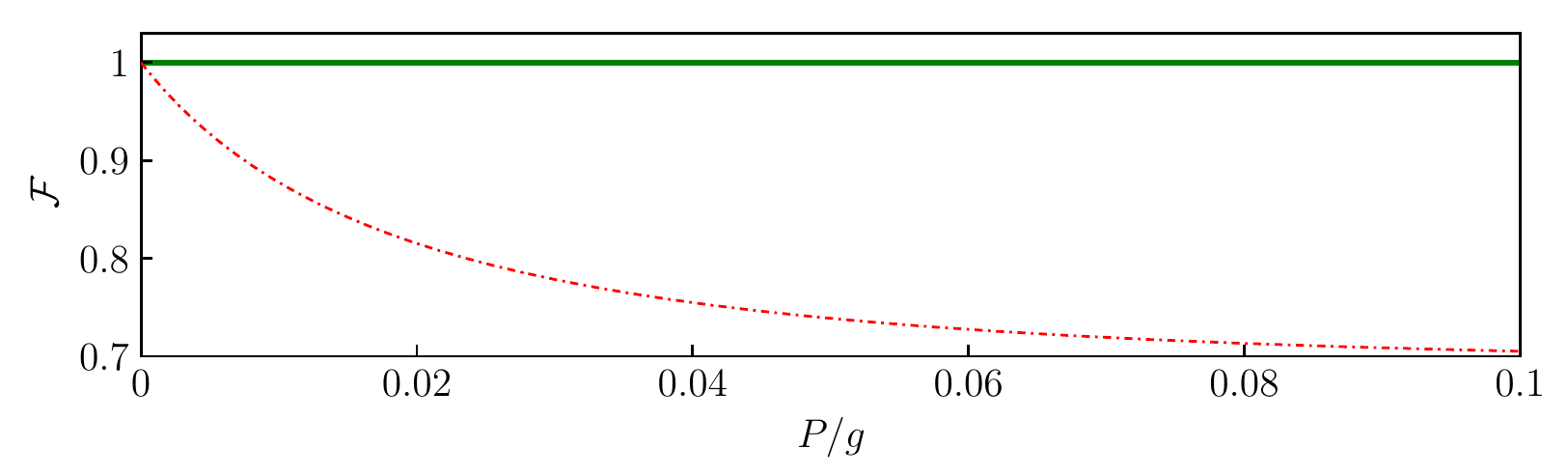}\caption{
Fidelity of the system steady-state as a function  of the pumping rate $P$, in particular, dot-dashed red curve  is the fidelity based on numerical calculations using the NHQM method and the exact Lindblad master equation approach. Solid green line corresponds to the fidelity between the corrected NHEH and the Lindblad master equation approaches.  (For interpretation of the references to color in this figure legend, the reader is referred to the web version of this article.)}\label{Fig2}
\end{figure}

\noindent
With the purpose of compare and investigate the performance of the two non-Hermitian approaches discussed previously, we have considered first the case where there is only dissipation in the JC model
and therefore the pumping rate of cavity photons is fixed to $P=0$. The other parameters of the system are: $\omega_x/g=1000$, $\omega_c/g=1000$, $\kappa/g=0.1$ and $\gamma_x/g=0.01$. In particular, we have solved numerically the non-linear dynamical equations given by Eq.~(\ref{bloch:eqs}) accordingly to the NHQM method, moreover, similar numerical calculations have been performed for both the corrected NHEH and the Lindblad master equation approaches given by Eqs.~(\ref{Fullcorrected}) and (\ref{exactnumEq}), respectively. Fig.~\ref{Fig1}(a) 
shows the Pearson's correlation coefficient \cite{Rodgers:1988} for both populations and coherences for a time evolution until a time $tg=1000$, and as a function of the parameter $\alpha$. It is important to clarify to the reader that, we have defined the initial conditions through the parameter $\alpha$, more precisely, $\rho_{G0G0}(0)=\alpha$ and $\rho_{G1G1}(0)=1-\alpha$ in the numerical solutions of each one of the theoretical approaches above mentioned. Since the correlation coefficient offers to measurement on how much different are two vectors, and in particular, its measurement is equal to one when they are the same. The correlation coefficient comparing the populations $\rho_{X0X0}$ (dot-dashed yellow line), $\rho_{G1G1}$ (solid thin blue line), $\rho_{G0G0}$ (dotted black line), and the coherences $\rho_{X0G1}$ (dashed red line) and $\rho_{G0G1}$ (double dot-dashed gray line) based on NHQM and the Lindblad master equation approaches are shown in panel (a). The numerical results for the correlation coefficient comparing both populations and coherences based on the Lindblad master equation and the corrected NHEH approaches are shown with a solid thick green line. Surprisingly, it is found that this correlation coefficient is equal to one for these numerical calculations whereas the NHQM method fails
to capture the full quantum dynamics, except for the particular initial condition $\rho_{G0G0}(0)\approx1$.
Fig.~\ref{Fig1}(b)-(c) shows the imaginary part of the coherences $Im[\rho_{X0G1}]$ and $Im[\rho_{G1G0}]$ as a function of time, notice that the numerical calculations based on Lindblad master equation, the corrected NHEH and NHQM approaches are displayed as solid black, dashed green and dot-dashed red lines, respectively. Similar plots are shown in Fig.~\ref{Fig1}(d)-(e)-(f), but now comparing the time evolution of the populations $\rho_{G0G0}$, $\rho_{G1G1}$ and $\rho_{X0X0}$ with the same color code as before. Clearly, it is observed that the NHQM method significantly deviates from the exact results and the foremost cause of the discrepancy is due to that, the usage of the normalized density operator introduces non-linear dynamics that overestimates the exact results even when the probabilistic interpretation of the quantum mechanics remains well-established. It is worth to mention that, in contrast to the NHQM method, the corrected NHEH approach guarantees the transference of population between adjacent rungs in the JC ladder and therefore the normalization condition remains valid for all times, i.e. $\tr(\hat{\rho})=1$. In what follows, we consider a more demanding situation for the non-Hermitian approaches mentioned in this paper. Specifically, we introduce the linear pumping of cavity photons at rate $P$ into the JC model through the anti-Hermitian operator $-iP\hat{a}\hat{a}^{\dagger}/2$ and therefore the anti-Hermitian part of the Eq.~(\ref{eq:001}) is now given by 
\begin{equation}\label{her:op:minus2}
\hat{H}_{-}=-i\frac{\kappa}{2}\hat{a}^{\dagger}\hat{a}-i\frac{\gamma_{x}}{2}\hat{\sigma}^{\dagger}\hat{\sigma}
-i\frac{P}{2}\hat{a}\hat{a}^{\dagger}.
\end{equation}
It is worth to mention that the NHQM method has initially been developed for describing dissipative processes, and to the best of our knowledge this is the first work that investigates the performance of this theoretical approach with gain through an anti-Hermitian operator. The presence of this term into the anti-Hermitian part provides us with a new set of non-linear dynamical equations similar to the equations given by Eq.~(\ref{bloch:eqs}), but they are not shown here. However, it is interesting to note that this new set of dynamical equations has more than one steady state when it is considered gain in the JC model, and there is no criterion within the NHQM method for choosing the right one. In fact, in our numerical simulations, we have found five possible steady states where four of them are solutions without a physical sense that must be rejected. On the other hand, the remaining solution coincides with the correct steady state predicted by the Eq.~(\ref{exactnumEq}) when the pumping of cavity photons vanishes. Taking into account this fact, we have performed numerical calculations of the fidelity that is defined by  $\mathcal{F}(\hat{\rho},\hat{\sigma})=\big[\tr(\sqrt{\sqrt{\hat{\rho}}\hat{\sigma}\sqrt{\hat{\rho}}})\big]^2$ as a measure of the distinguishability between two arbitrary density operators $\hat{\rho}$ and $\hat{\sigma}$ \cite{Jozsa:1994}. By definition $\mathcal{F}(\hat{\rho},\hat{\sigma})=1$ if and only if $\hat{\rho}=\hat{\sigma}$, that corresponds to the maximum value of the fidelity when the two density operators are completely indistinguishable. Fig.~\ref{Fig2}
shows numerically-computed fidelity of the system steady-state as a function of the pumping rate $P$. Dot-dashed red line is the fidelity numerically-obtained steady state based on the NHQM method (see Eqs.~(\ref{eq:004}), (\ref{her:op:plus}) and (\ref{her:op:minus2}) and the Lindblad master equation approach (see Eq.~(\ref{exactnumEq})). The solid green line corresponds to the fidelity numerically-obtained steady state based on the corrected NHEH and the Lindblad master equation approaches. It is found that the NHQM method fails for modeling open quantum dynamics with any source of gain in the system, since the steady state predicted by the NHQM method is always the solution of the dissipative dynamics such that $\rho_{G0G0}=1$. On the other hand, the corrected NHEH approach is well suited to describe quantum dynamics involving simultaneously dissipation and gain in the system, since the fidelity remains equal to one. 
\section{Conclusions}\label{conclusions}
\noindent
The findings of this study indicate that the NHQM method is not well suited to describe open quantum dynamics, and in general this methodology fails for arbitrary initial conditions as well as for describing the correct steady-state of the system when are considered pumping conditions. We have confirmed numerically that the corrected NHEH approach leads to results that are in excellent agreement with those obtained by the Lindblad master equation formalism. Moreover, this theoretical approach works properly and it can be used for studying more general situations where simultaneously appears dissipation and gain in the system, in contrast to the NHQM method that can be considered only for dissipative dynamics. Finally, we are confident that our results may improve knowledge about the energy transfer in quantum photobiological complexes where the NHQM method has been previously applied~\cite{Zloshchastiev:2017}.
\section*{Acknowledgments}
\noindent
We would like to thank K. G. Zloshchastiev for calling our attention to the fact that the NHQM method forces to introduce a normalized density operator. S.E.-A. and H.V.-P. gratefully acknowledge funding by
COLCIENCIAS projects  ``Emisi\'on en sistemas de Qubits
Superconductores acoplados a la radiaci\'on'', c\'odigo
110171249692, CT 293-2016, HERMES 31361, ``Control din\'amico de la emisi\'on en sistemas de qubits acoplados con cavidades no-estacionarias'', c\'odigo 201010028998, HERMES 41611 and ``Interacci\' on radiaci\'on-materia mediada por fonones en la electrodin\'amica cu\'antica de cavidades'', c\'odigo
201010028651, HERMES 42134. S.E.-A. also acknowledges support from the  ``Beca de
Doctorados Nacionales'' de COLCIENCIAS convocatoria 727. E.A.G acknowledges financial support from
Vicerrector\'ia de Investigaciones of the Universidad del Quind\'io through the project No. 919.

\end{document}